\documentclass{webofc}
\usepackage[varg]{txfonts}   
\usepackage{graphicx}
\usepackage{subcaption}
\usepackage{lineno}
\newcommand{\pt}{p_T}
\newcommand{\chisqip}{\chi^2_{\textrm{IP}}}
\newcommand{\chisqvtx}{\chi^2_{\textrm{vtx}}}
\newcommand{\gev}{\textrm{GeV}}
\newcommand{\mev}{\textrm{MeV}}

\begin{document}
%
\title{Applications of Lipschitz neural networks to the Run 3 LHCb trigger system}
%
%

\author{\firstname{Blaise} \lastname{Delaney}\inst{1,2}\fnsep\thanks{\email{blaise.delaney@cern.ch}} \and
        \firstname{Nicole} \lastname{Schulte}\inst{3} \and
        \firstname{Gregory} \lastname{Ciezarek}\inst{4} \and
        \firstname{Niklas} \lastname{Nolte}\inst{5} \and
        \firstname{Mike} \lastname{Williams}\inst{1,2} \and
        \firstname{Johannes} \lastname{Albrecht}\inst{3}
}

\institute{Massachusetts Institute of Technology, Cambridge MA, USA
\and
    NSF AI Institute for Artificial Intelligence and Fundamental Interactions (IAIFI) 
\and
    TU Dortmund University, Dortmund, Germany
\and 
    CERN, Meyrin, Switzerland
\and 
            Meta AI (FAIR)
}



\abstract{%
The operating conditions defining the current data taking campaign at the Large Hadron Collider, known as Run 3, present unparalleled challenges for the real-time data acquisition workflow of the LHCb experiment at CERN. To address the anticipated surge in luminosity and consequent event rate, the LHCb experiment is transitioning to a fully software-based trigger system.
This evolution necessitated innovations in hardware configurations, software
paradigms, and algorithmic design. A significant advancement is the integration of monotonic Lipschitz neural networks into the LHCb trigger system. 
These deep learning models offer certified robustness against detector instabilities, and the ability to encode domain-specific inductive biases.
Such properties are crucial for the inclusive heavy-flavour triggers and, most notably, for the topological triggers designed to inclusively select $b$-hadron candidates by exploiting the unique kinematic and decay topologies of beauty decays. This paper describes the recent progress in integrating Lipschitz neural networks into the topological triggers, highlighting the resulting enhanced sensitivity to highly displaced multi-body candidates produced within the LHCb acceptance.
}

\maketitle
%






\section{Introduction}

The LHCb detector~\cite{LHCb:detector} located at the Large Hadron Collider at CERN is a single-arm forward spectrometer instrumented to achieve acceptance in the pseudorapidity range, $2 < \eta < 5$. 
The primary goal of the LHCb experiment is the discovery of Beyond the Standard Model (BSM) physics through the analysis of heavy-flavour processes, with particular focus placed on investigating $b$-hadron decays. Since its inception, however, the LHCb physics programme has grown substantially to include, among other endeavours, the search for feebly interacting dark-portal candidates produced in the LHCb geometrical acceptance~\cite{LHCb:dark-photon-i, LHCb:dark-photon-ii, LHCb:lowmass_mumu, LHCb:dark-boson-i, LHCb:dark-boson-ii}. 

Throughout its Run 3 data taking campaign, the LHCb experiment is tasked with operating under unprecedented conditions, namely a nominal instantaneous luminosity, \mbox{$\mathcal{L} = 2 \times 10^{33}~\textrm{cm}^{-2}\textrm{s}^{-1}$}, amounting to a five-fold increase on the data taking conditions present in Runs 1 and 2. In order to cope with the challenging detector occupancy of Run 3, the LHCb experiment has pioneered the deployment of a redesigned, fully software-based trigger system for real-time data acquisition. In this paradigm, the full detector readout and event building is enacted at an incoming 30 MHz rate of visible proton-proton ($pp$) bunch crossings. A two-staged trigger system is deployed to select events of interest with a 100 kHz output rate suitable for storage.

\begin{figure}[t]
\centering
\includegraphics[width=\textwidth]{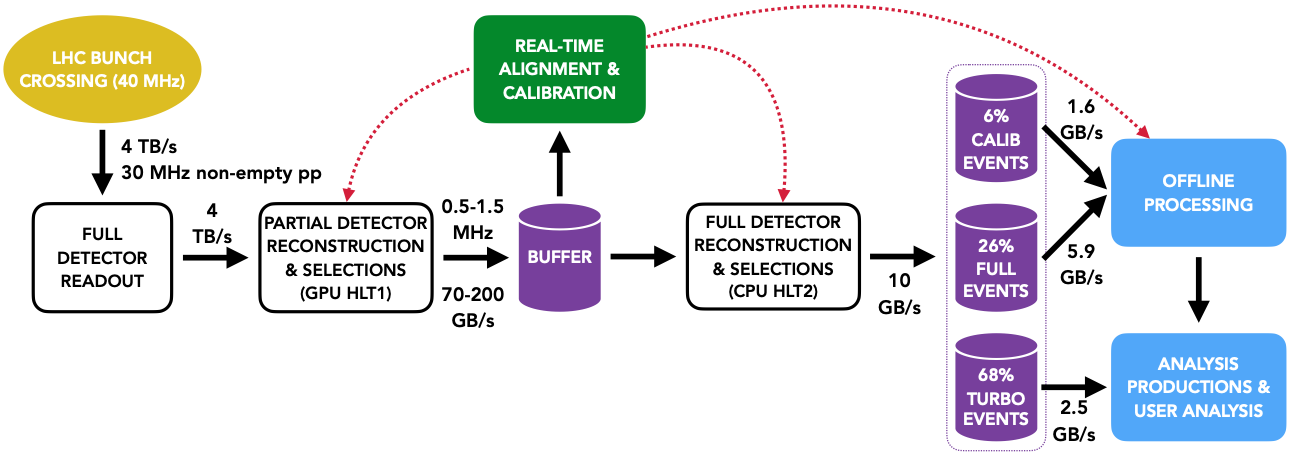}
\caption{Schematic representation of the LHCb data flow during the Run 3 data taking period. The incoming rate of 30 MHz of non-empty bunch crossings is processed by the full detector readout and thereafter passed through the two-tiered trigger system. The selected data is stored and further processed offline for end-user analysis. Taken from~\cite{LHCB-FIGURE-2020-016}.}
\label{fig:lhcb-data-flow}       
\end{figure}

The core objective of the LHCb trigger is therefore data-volume reduction within the real-time data storage constraints. This process translates to a reduction by a factor of approximately 400 of the input 4 TB/s bandwidth, at nominal instantaneous luminosity, to achieve an output bandwidth of 10 GB/s~\cite{LHCb:UpgradeI}.
Figure~\ref{fig:lhcb-data-flow} provides a schematic representation of the LHCb data flow in Run 3. Following the full detector readout, the GPU-enabled first trigger stage, the High Level Trigger 1 (HLT1), delivers partial event reconstruction from charged-track information, resulting in a data-volume reduction by a factor of approximately 20. Subsequently, the data is passed to a buffer stage to achieve real-time alignment and calibration, thus enabling the CPU-based High Level Trigger 2 (HLT2). This trigger stage operates selection algorithms exploiting offline-quality, full-event reconstruction observables.

To achieve the real-time rate reduction required by the Run 3 operating conditions, the LHCb trigger exploits a combination of expert systems and machine learning solutions. Lipschitz monotonic neural networks (NNs) find optimal application in the latter, specifically in the inclusive algorithms deployed to select heavy-flavour particles.

\section{Applications of Lipschitz neural networks to the LHCb trigger}
Implementing trigger selections is a critical step in High Energy Physics data acquisition, as discarded events are irreversibly lost. No margin for error is therefore afforded to the event-selection protocols implemented in the trigger system. Moreover, effective classification algorithms must be able to process event-level information within the memory and compute limits imposed by real-time data acquisition. The application of monotonic Lipschitz NNs to the LHCb trigger meets such mission-critical requirements whilst delivering additional desirable certified benefits: namely, robustness and interpretability.

In this work, robustness signifies mitigated sensitivity to both simulation inaccuracies and detector instabilities during data taking. Such conditions may be achieved by constraining the gradient of the function approximated by a deep learning classifier with respect to the input features. The models deployed in the LHCb experiment realise this result by bounding the Lipschitz constant of the learnt decision-boundary function. This approach effectively sets a strict upper limit on the classifier-response variation resulting from experimental instabilities of limited magnitude.
Robustness is thus essential for the estimators deployed in the trigger system, serving to reduce sensitivity to resolution and calibration effects conditioning the input features in real time. Additionally, robustness simplifies the evaluation of trigger-related systematic uncertainties in end-user physics measurements.

A complementary advantage is offered by enforcing a monotonic estimator response with respect to a set of input features. In essence, this procedure expresses domain-specific inductive bias. 
Deploying provably monotonic networks in the trigger system enhances the selective retention of interesting outlier candidates absent from training samples and, therefore, not learnt by the trigger models.
Specifically, in the context of the LHCb heavy-flavour triggers, such architectures enable enhanced sensitivity to potential yet-undiscovered feebly interacting BSM states produced within the LHCb acceptance.


By adopting the deep learning models introduced in ref.~\cite{Kitouni:2021fkh}, robustness and certified monotonic response with respect to a set of input features are realised through a minimal set of architecture-level constraints. Such conditions enable highly expressive classifiers capable of inference within the bandwidth-rate constraints set by the LHCb Run 3 trigger operations. Consequently, Lipschitz monotonic NNs supersede the decision-tree implementations of the inclusive heavy-flavour triggers deployed in previous data taking campaigns~\cite{Gligorov:2012qt, Likhomanenko:2015aba}. The topological triggers, presented in the following section, exemplify the advantages obtained through this methodological advancement.

\section{The LHCb topological triggers}
The topological triggers are designed to inclusively select $b$-hadron decay processes based on the offline-quality information available at the HLT2 trigger stage. Decays of beauty particles display a distinct topology: owing to their forward boost in the detector, they exhibit a long lifetime~\cite{Workman:2022ynf}, causing them to traverse distances of $\mathcal{O}(1\,\textrm{cm})$ before decaying within the detector acceptance. Additionally, owing to their relatively high mass, beauty hadrons typically exhibit sizeable transverse momentum, $p_{T}$.
Combined, both conditions make for a distinct experimental signature. The topological triggers thus aim to select displaced secondary decay-vertex candidates reconstructed in the pairwise combination of final-state charged tracks. 

Two variations of the Run 3 topological architectures have been incorporated in the LHCb online software stack, henceforth referred to as the two- and three-body topological triggers. 
These are designed to identify beauty secondary-vertex candidates reconstructed in the combination of two or three charged particles, respectively. Such a suite of topological triggers thus facilitates the identification of multi-particle beauty decays.
The two-body composites are reconstructed from final-state tracks compatible with originating from a well-resolved secondary-vertex candidate appreciably displaced from the $pp$ collision point.
The three-body candidates, in turn, are reconstructed by adding a companion track to the two-body object and imposing similar vertex-fit and kinematic criteria as for the two-body counterparts. 
Notably, the topological triggers record the full-event level information, writing to the so-called \emph{Full Stream}, depicted in Figure~\ref{fig:lhcb-data-flow}.
As a result, $n$-body $b$-hadron decays, with $n>3$, may be successfully reconstructed by combining additional tracks persisted in the event with the trigger-selected two- and three-body candidates. Crucially, this design maximises the selection efficiency on signal whilst meeting the HLT2 output-rate requirements.

The HLT2 topological triggers are tasked with rejecting several sources of background: 
combinatorial and soft-QCD processes; 
interactions of particles with the LHCb detector material, exhibiting softer $p_T$ than the beauty signal, and a comparatively larger flight distance from the $pp$ interaction point; 
fake particles, typically referred to as \emph{ghosts}, erroneously inferred from tracking-level information and exhibiting high $p_T$, as a linear trajectory within the detector translates to the maximal possible momentum reconstructed by the tracking system;
and charm decays, rendered challenging by the high charm production cross section, approximately $\mathcal{O}(10)$ times higher than the beauty counterpart, and topologies compatible with the signal, albeit with comparatively reduced lifetimes~\cite{Workman:2022ynf}.

Each topological trigger exploits a two-staged selection to achieve effective signal isolation. 
Firstly, a cut-based prefilter is devised to discard prompt and soft background candidates. 
Subsequently, a deep learning classifier is implemented as a monotonic Lipschitz neural network.
The two- and three-body models share the same architecture complexity, amounting to four hidden layers comprising 16, 32, 64, and 32 internal nodes, respectively. This design choice mitigates inference-time consumption during real-time trigger operations. 
Model training is enacted through an offline Python-based pipeline exploiting the \textsc{PyTorch}~\cite{paszke2019pytorch} software package. Thereafter, the network weights are ported to the LHCb trigger software stack for event-based inference in real time.  

\begin{table}[t]
\centering
\caption{Features used to train the two-body classifier, along with the monotonicity requirements and the operations enacted to rescale the feature distributions to a range of order unity. The operators $\cdot\{\}$ and $\sum$ run over the final-state tracks, and are evaluated on a per-candidate basis. The shorthand notations $\gev$ and $\log$ signify rescaling from units of $\mev/c^{(2)}$ to $\gev/c^{(2)}$ and evaluating the natural logarithm of the observable, respectively.}
\label{tab:2body_features}
\begin{tabular}{lcc}
\hline
Feature & Monotonicity & Rescaling to $\mathcal{O}(1)$ procedure \\
\hline
$\min\{\pt\}$ & Increasing & \gev \\
$\sum{\pt}$ & Increasing & \gev \\
$\min\{\chisqip\}$ & Increasing & $\log$ \\
$\max\{\chisqip\}$ & --- & $\log$ \\
Composite $\pt$ & Increasing & \gev \\
Composite $\max({\textrm{DOCA}})$ & --- & --- \\
Composite $m_\textrm{corr}$ & --- & \gev \\
Composite flight distance $\chi^2$ & --- & $\log$ \\
Composite $\chisqvtx$ & --- & --- \\\hline
\end{tabular}

\label{tab:twobody_features}
\end{table}


Both topological triggers are optimised by considering a suite of exclusive simulations representative of the LHCb beauty physics programme. In this way, the respective selection criteria and architecture complexity are optimised to attain sensitivity to the characteristic decay topologies and kinematics of $b$-hadron decays.
The exclusive signal Monte Carlo (MC) simulations are combined to contribute the same number of decay-vertex candidates 
to the inclusive signal sample. This, in turn, is used both to optimise the cut-based prefilter and train the NNs. 
Crucially, this procedure prevents potential biases in the form of heightened sensitivity to a subset of signal channels, by construction. 
The background is modelled by an inclusive minimum-bias MC sample, generated to represent the average content of a $pp$ collision. Notably, events containing beauty candidates are filtered from the minimum-bias sample, making it a suitable inclusive proxy of the aforementioned background processes.

\begin{table}[ht]
\centering
\caption{Features used to train the three-body classifier, along with the monotonicity requirements and the operations enacted to rescale the feature distributions to a range of order unity. The operators $\cdot\{\}$ and $\sum$ run over all final-state tracks, and are evaluated on a per-candidate basis. Conversely, the $\cdot_{\textrm{2body}}\{\}$ operations run over the two-body children only. The shorthand notations $\gev$ and $\log$ signify rescaling from units of $\mev/c^{(2)}$ to $\gev/c^{(2)}$ and evaluating the natural logarithm of the observable, respectively.}
\label{tab:3body_features}
\begin{tabular}{lcc}
\hline
{Feature} & {Monotonicity} & {Rescaling to $\mathcal{O}(1)$ procedure} \\
\hline
$\min\{\pt\}$ & Increasing & \gev \\
$\sum{\pt}$ & Increasing & \gev \\
$\min_{\textrm{2body}}\{\pt\}$ & Increasing & \gev \\
$\sum_{\textrm{2body}}{\pt}$ & Increasing & \gev \\
$\min\{\chisqip\}$ & Increasing & $\log$ \\
$\max\{\chisqip\}$ & --- & $\log$ \\
Three-body $\pt$ & Increasing & \gev \\
Two-body child $\pt$ & Increasing & \gev \\
Three-body $\max({\textrm{DOCA}})$ & --- & --- \\
Two-body child $\max({\textrm{DOCA}})$ & --- & --- \\
Two-body child $\chisqip$ & --- & $\log$ \\
Two-body child $m_\textrm{corr}$ & --- & \gev \\
Three-body  $m_\textrm{corr}$ & --- & \gev \\
Three-body flight distance $\chi^2$ & --- & $\log$ \\
Two-body child flight distance $\chi^2$ & --- & $\log$ \\
Three-body $\chisqvtx$ & --- & $\log$ \\
Two-body child $\chisqvtx$ & --- & --- \\
\hline
\end{tabular}
\label{tab:threebody_features}
\end{table}
The NNs are trained on feature sets maximising the post-prefilter signal-to-background divergence whilst maintaining low pairwise feature correlation. The two- and three-body feature sets are presented in Tables~\ref{tab:2body_features} and~\ref{tab:3body_features}. Crucially, the adopted inputs must capture the decay topology and kinematics of multi-body beauty composite candidates. To this end, the NNs are trained on a suite of kinematic and geometric observables comprising the candidate transverse momentum, $p_T$; the decay-vertex fit quality, $\chi^2_{\textrm{vtx}}$; flight distance $\chi^2$; the charged-tracks distance of closest approach (DOCA); the multi-body corrected mass value, $m_{\textrm{corr}}$~\cite{mcorr}; and the IP $\chi^2$, the impact parameter $\chi^2$ with respect to the primary vertex of the multi-body decay-vertex objects. Similar kinematic and geometric features are extracted from the charged final-state tracks. 

The features are subjected to a preprocessing stage constraining each input observable distribution to a range of $\mathcal{O}(1)$. 
Additionally, a $5\sigma$ window is imposed about the mean of each feature distribution. Input values exceeding the retention boundaries are mapped onto the last value accepted in the per-feature selection interval.
Combined, these preprocessing steps stabilise the performance of the classifier without depleting the statistical population of the training samples. Moreover, comparing similarly ranged distributions facilitates the choice of the Lipschitz-constant bound, $\lambda$, of the learnt decision functions.

Separate upper bounds are imposed on the two- and three-body Lipschitz constants, respectively. These are individually selected to maximise reconstruction efficiency on the multi-body signal candidates whilst maintaining compatibility of the resolution expected of the LHCb detector. In this way, the Lipschitz-bound assignments prevent significant variations in the classification score produced by the estimators, in the limit of the inputs varying within their respective resolution scale. 

\begin{figure}
    \centering
    \begin{subfigure}{0.66\textwidth}
        \centering
        \includegraphics[width=\linewidth]{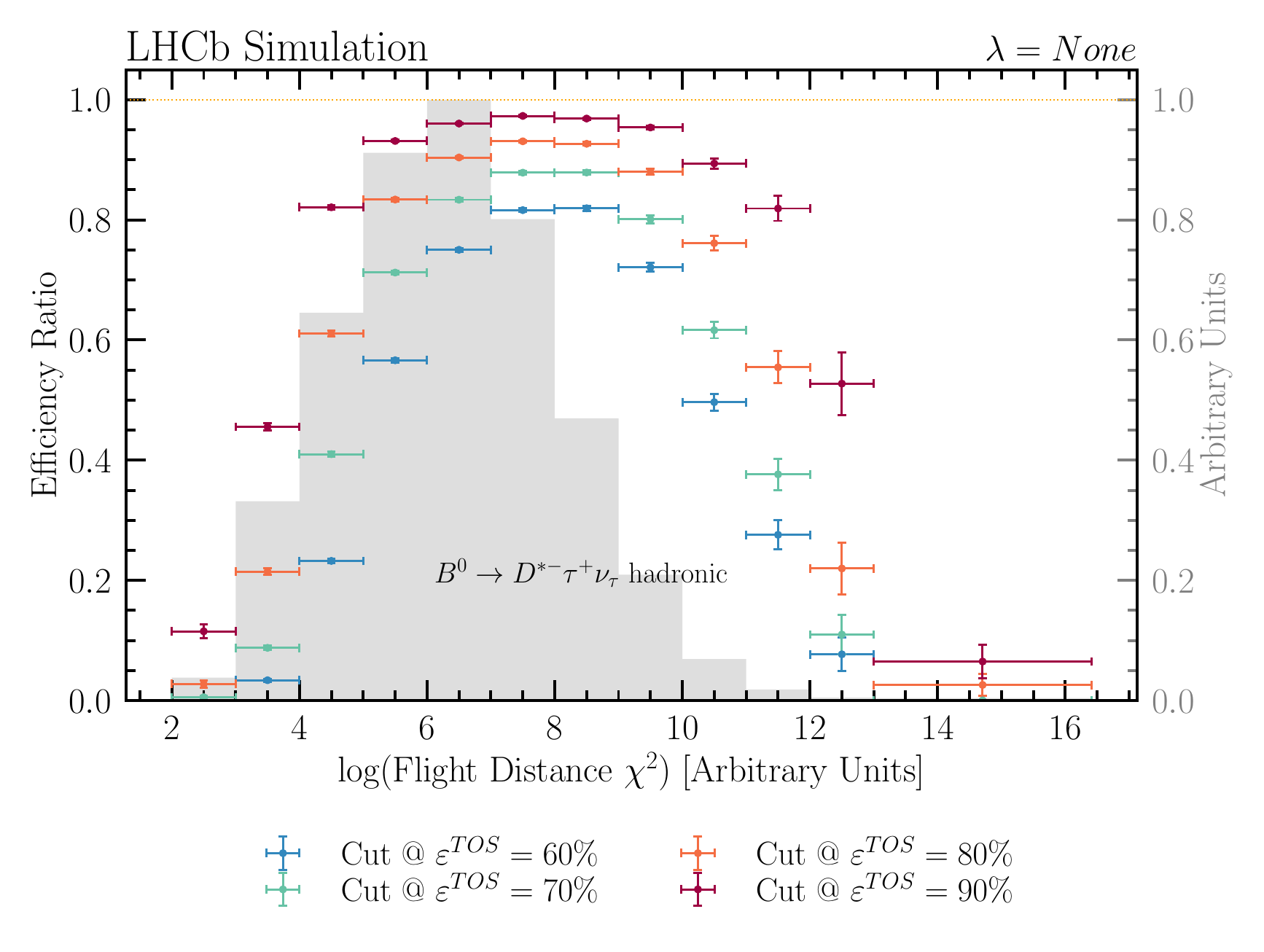} 
        \label{fig:sub1}
    \end{subfigure}
    \begin{subfigure}{0.66\textwidth}
        \centering
        \includegraphics[width=\linewidth]{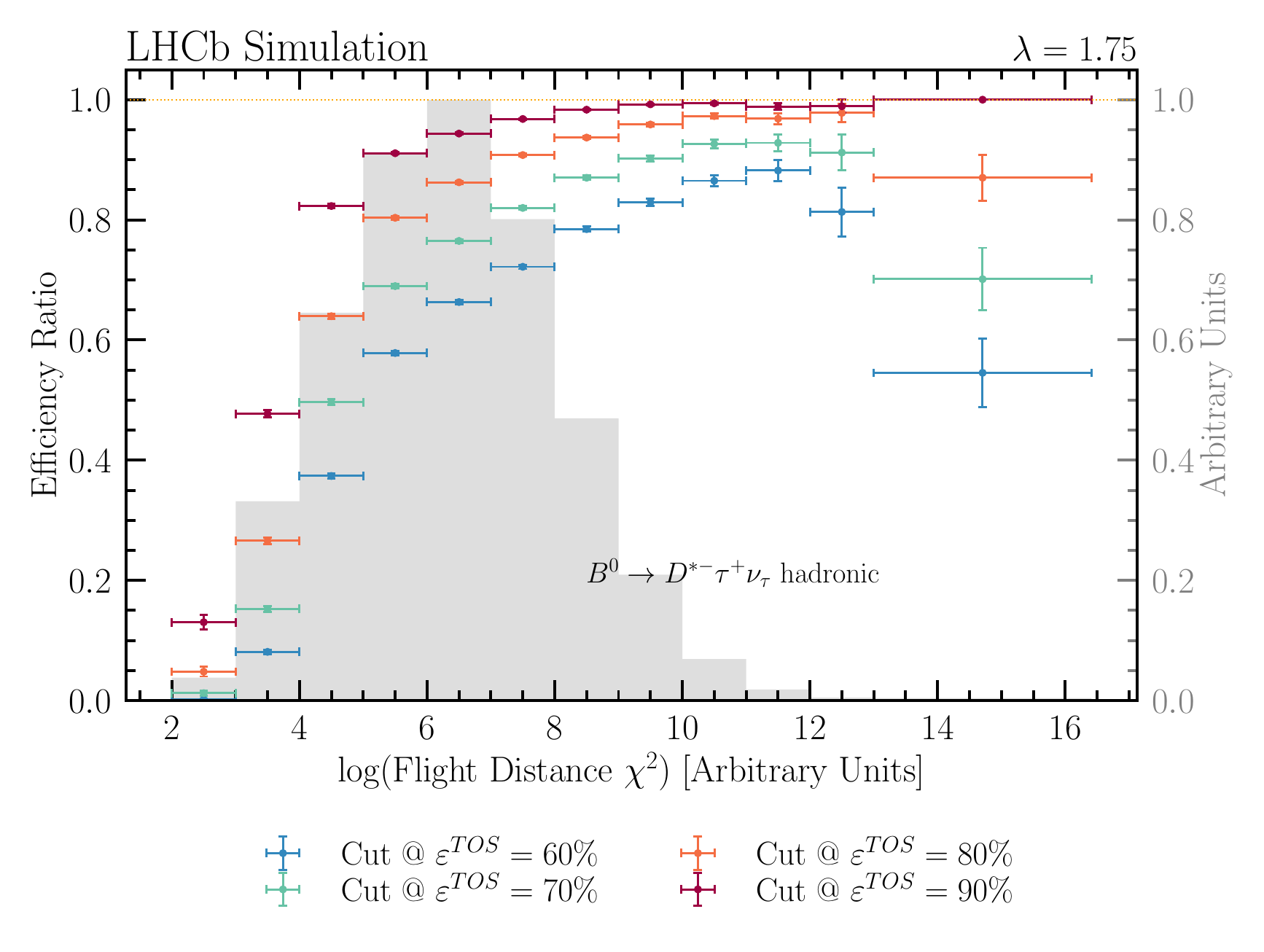} 
        \label{fig:sub2}
    \end{subfigure}
    \caption{Selection Efficiency on simulated $B^0 \to D^{*-}\tau^+\nu_{\tau}$ candidates reconstructed from two final-state tracks. Top: performance delivered by a feed-forward, unconstrained network. Bottom: efficiency values obtained by the two-body monotonic Lipschitz network.
    The normalised $B^0$ logarithmic flight distance $\chi^2$ distribution is shown in the shaded grey histogram. 
    Binwise efficiencies for response cuts yielding global trigger-on-signal efficiency values, $\varepsilon^{TOS}$, at the $60\%$-$90\%$ level are marked. The Lipschitz bound, $\lambda$, is shown where relevant.
    }
    \label{fig:tos-results}
\end{figure}

Finally, a monotonically increasing response is enforced with respect to a subset of features, identified independently for the two- and three-body classifiers. Monotonicity is imposed in the IP $\chi^2$ and the $p_T$ budget of the composite candidates and their constituents. This conservative choice 
prevents the introduction of unwanted biases in selection efficiency due to monotonicity.
Furthermore, this approach delivers enhanced sensitivity to highly displaced outliers compared to a baseline unconstrained model with identical network complexity, as demonstrated by Figure~\ref{fig:tos-results}. 
The per-candidate selection efficiency delivered by the two-body topological trigger is evaluated for simulated $B^0 \to D^{*-}\tau^+\nu_{\tau}$ decays, where the $\tau$ lepton decays hadronically, excluded by the training MC samples. 
The efficiency extraction is performed in bins of the signal logarithmic $\chi^2$ fit-quality of the partially reconstructed $B^0$ candidate flight distance with respect to the $pp$ collision point, taken to be a proxy for the decay-vertex candidate lifetime.
Compared to a performance attained by the baseline unconstrained NN, the topological trigger delivers comparatively higher selection efficiencies for highly displaced outliers. Concurrently, it delivers compatible performance in the remainder of the observed range. 
Such a trend is evident when requiring thresholds on the respective network responses delivering overall efficiency values at the $60\%$-, $70\%$-, $80\%$-, and $90\%$-level on signal. The efficiency drop evident at very high displacement is due to the fact that monotonicity is enforced in a subset of the input features. Consequently, all else being equal, the distributions of other features can be much more background-like at large flight distance, leading to a marginal efficiency drop in the high tail of the flight distance $\chi^2$ projection.
Nevertheless, the results presented in Figure~\ref{fig:tos-results} demonstrate the capacity of the two-body topological trigger to efficiently select signal $b$-hadron decays and enhance in sensitivity to highly displaced candidates. 
This property bolsters the capacity to retain outlier multi-body objects, and thus feebly interacting BSM candidates produced in the LHCb acceptance.

\section{Summary and Discussion}
LHCb is pioneering the deployment of a fully software-based trigger in the LHC Run 3 data taking campaign. Such a paradigm shift facilitates the adoption of Lipschitz monotonic NNs in the LHCb inclusive heavy-flavour triggers. These models are particularly well suited to the real-time selection of events containing decay-vertex candidates exhibiting topologies and kinematics compatible with heavy-flavour decays.

The utility of monotonic Lipschitz NNs is exemplified by the topological triggers deployed in the highest level of the LHCb Run 3 trigger system, HLT2.
The preliminary results presented in this contribution demonstrate capacity to efficiently select two- and three-body signal candidates. 
Enforcing a monotonically increasing response with respect to a subset of input features yields enhanced sensitivity to highly displaced decay-vertex candidates, as evaluated on a probe decay channel.
From a physics standpoint, this performance strengthens sensitivity to beauty candidates at high lifetime and, notably, to potential BSM states produced within the LHCb detector. Furthermore, the Lipschitz bound applied to the learnt decision function provides certified protection against instabilities during data taking, thereby easing the evaluation of the relevant systematic uncertainties in end-user measurements.

The combined output of the topological triggers dominates the HLT2 bandwidth allocation. The threshold imposed on the classifier response for real-time inference on data, in turn, is fixed by the maximum bandwidth allocated to the HLT2 trigger stage when writing data to storage. 
Optimization studies are currently underway to improve the classification power of these triggers. The aim is to maximise selection efficiency for standard-candle beauty decay modes whilst satisfying the HLT2 output-rate constraints.

Finally, as robustness and monotonicity are ideal inductive biases in experimental particle physics, investigations are progressing to deploy Lipschitz NNs in the Run 3 LHCb tracking, lepton identification and ghost-rejection algorithms.

\section{Acknowledgements}

The authors would like to thank the LHCb computing and simulation teams for their support and for producing the simulated LHCb samples used in the paper. The authors would also like to thank the LHCb online, DPA and RTA team for providing most of the software that this project was built upon and for all the support and guidance that was given. BD and MW were supported by NSF grant PHY-2019786 (The NSF AI Institute for Artificial Intelligence and Fundamental Interactions, http://iaifi.org/). BD and MW were also supported by U.S. NSF grant PHY-2209181. NS and JA acknowledge funding from the German Science Foundation DFG, within the Collaborative Research Center SFB1491 ``Cosmic Interacting Matters - From Source to Signal.'' JA acknowledges funding from the German Federal Ministry of Education and Research (BMBF, grant no. 05H21PECL1) within ErUM-FSP T04.

%
\bibliography{references}

\end{document}